\documentclass[aps,prl,superscriptaddress,showpacs,twocolumn]{revtex4-1}

\usepackage[utf8]{inputenc}
\usepackage[english]{babel}
\usepackage{times}
\usepackage{color}
\usepackage{graphicx}
\usepackage{import}

\begin{document}
\title{Quantum versus Classical Annealing of Ising Spin Glasses}
\author{Bettina Heim}
\affiliation{Theoretische Physik, ETH Zurich, 8093 Zurich, Switzerland}
\author{Troels F. R{\o}nnow}
\affiliation{Theoretische Physik, ETH Zurich, 8093 Zurich, Switzerland}
\author{Sergei V. Isakov}
\affiliation{Google, Brandschenkestrasse 110, 8002 Zurich, Switzerland}
\author{Matthias Troyer$^*$}
\affiliation{Theoretische Physik, ETH Zurich, 8093 Zurich, Switzerland}
\begin{abstract}
The strongest evidence for superiority of quantum annealing on spin glass problems has come from comparing simulated quantum annealing using quantum Monte Carlo (QMC) methods to simulated classical annealing [G. Santoro {\it et al.}, Science 295, 2427(2002)]. Motivated by experiments on programmable quantum annealing devices we revisit the question of when quantum speedup may be expected for Ising spin glass problems.  We find that even though a better scaling compared to simulated classical annealing can be achieved for QMC simulations, this  advantage is due to time discretization and measurements which are not possible on a physical quantum annealing device. QMC simulations in the physically relevant continuous time limit, on the other hand, do not show superiority. Our results imply that care has to be taken when using QMC simulations to assess quantum speedup potential and are consistent with recent arguments that no quantum speedup should be expected for two-dimensional spin glass problems.
\end{abstract}
\maketitle

With  first archeological records dating back more than six thousand years \cite{annealing_first}, thermal annealing is likely
to be the oldest optimization method in human history. By first heating a material and then
letting it cool down slowly, it can relieve internal
stresses and achieve a lower  energy
state. Inspired by thermal annealing, Kirkpatrick and co-authors
suggested a similar approach to find the ground states of combinatorial optimization problems
 more than three decades ago \cite{kirkpatrick1983}. In particular, they studied Ising spin glasses with $N$
spins described by the Hamiltonian
\begin{equation}
\label{eq:classical}
H_c = - \sum_{i<j} J_{ij}s_is_j - \sum_{i} h_{i}s_i,
\end{equation}
where $s_i$ takes the values $\pm 1$ and represents the orientation of the
spin on lattice site $i$. The couplings between
spin $i$ and $j$ are denoted by $J_{ij}$ and $h_i$ are local fields. 

Non-convex optimization problems, such as finding the ground state of this Ising spin glass \cite{Barahona1982}, are important in many areas of science
and industry. Other typical problems include job scheduling \cite{jobschedule}, circuit
minimization \cite{Knuth}, and chain optimization \cite{scan_chain_opt} which are all
non-deterministic polynomially (NP) hard problems \cite{Cook-ACM-1971}. A consequence of
NP-hardness is that there exists a polynomial time mapping from one
problem to the other. Thus, any method to efficiently find
solutions to the Ising spin glass problem would provide an efficient way of solving other important problems.

Applying the Metropolis algorithm \cite{Metropolis}, Kirkpatrick {\em
et al.}
demonstrated that using ``simulated annealing'' (SA) --  simulating the process of cooling Ising spin glasses -- 
is an excellent method to minimize $H_c$. Starting from a high temperature where the system thermalizes quickly, the temperature is slowly decreased towards
zero. Thermal excitations allow the system to escape from local minima and relax into a low-energy state with energy equal or close
to that of the ground state $E_0$ \cite{sa_and_sa_conv_cond}. We will refer to the difference between 
the final energy $E$ and $E_0$ as the residual energy $E_{\rm res}=E-E_{0}$.

Quantum annealing (QA) \cite{Ray1989,Finnila1994,Kadowaki1998,idea_of_qa,qareview} uses a similar idea but employs quantum tunneling instead of thermal excitations to escape from local minima. QA can be advantageous in systems with narrow but tall barriers, which are easier to tunnel through than to thermally climb over.
To perform QA of Ising spin glasses, an additional
non-commuting kinetic term is added, usually by applying a transverse
magnetic field.
The time-dependent Hamiltonian of QA is then given by
\begin{equation}
\label{eq:quantum}
H_q = -\sum\limits_{ i<j} J_{ij} \sigma_i^z\sigma_j^z - \sum_{i} h_{i}\sigma^z_i - \Gamma(t) \sum\limits_i\sigma_i^x
\end{equation}
where $\sigma_i^z$ and $\sigma_i^x$ are Pauli $z$- and $x$-operators, respectively.
The transverse field $\Gamma(t)$ is initially much larger then the couplings, $\Gamma(0) \gg |J_{ij}|, |h_i|$, and the spins start out aligned in the $x$-direction. During quantum annealing $\Gamma(t)$ is slowly reduced to zero such that at the end of the
annealing process we recover the Hamiltonian of the initial Ising spin glass problem.
On a perfectly coherent quantum device, 
this algorithm \cite{idea_of_qa} would find the ground state of
the spin glass in question with probability approaching unity, provided that the annealing time $t_a$ is
sufficiently long to stay adiabatically in the ground state \cite{landau,zener}. 
Quantum annealing can also be performed at non-zero temperature, for example on spin glass material \cite{Brooke1999} or in programmable devices by the Canadian company D-Wave systems \cite{Johnson2011}.

\begin{figure}
  \centering  
  \def\svgwidth{\columnwidth}
  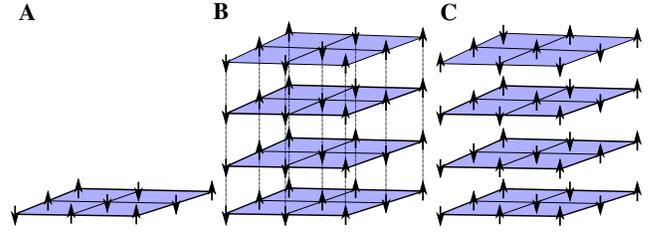
  \caption{A) Sketch of a small square lattice of classical spins used in SA. B) Path-integral QMC is performed after mapping the transverse field Ising model on the same lattie to a classical system representing imaginary time paths of the quantum spins in an additional imaginary time direction. We show an example of $M=4$ time slices with discrete time steps $\Delta_\tau = \beta/M$. C) With similar computational effort we can perform SA on $M$ independent replicas.}
\label{fig:path}
\end{figure}

QA can also be implemented as a simulation on a classical computer. While the simulation of unitary time evolution scales exponentially with the system size, QA can be efficiently performed using stochastic dynamics in a path integral quantum Monte Carlo (QMC) simulation \cite{santoro1,santoro2}. There, the partition function of the Ising model in a transverse field is mapped to that of a classical Ising model in one higher dimension corresponding to the imaginary time direction \cite{suzuki_orig}. We call  this algorithm simulated quantum annealing (SQA). In Fig. \ref{fig:path}A the two-dimensional lattice on which we perform SA and in Fig. \ref{fig:path}B the three-dimensional lattice on which SQA is performed after the path-integral mapping. Details  of the simulation algorithms and annealing schedules used in the Letter are discussed in the Supplementary Material.

 The strongest evidence for quantum annealing being superior to classical annealing for Ising spin glass instances comes from a comparison of the performance of SQA and SA \cite{santoro2,santoro1,qareview}. Upon increasing the annealing time the residual energy was seen to drop faster in SQA than in SA, indicating that quantum tunneling may indeed be advantageous in finding low energy states. However, recent studies of the performance of the D-Wave devices failed to see indications of quantum speedup \cite{article_eth}, although the device performance was consistent with that of a quantum annealer \cite{Boixo:2014ej}. Furthermore, in contrast to Refs. \cite{santoro2,santoro1} no advantage of SQA over SA was seen.

\begin{figure}[t]
  \centering  
  \includegraphics[width=\columnwidth]{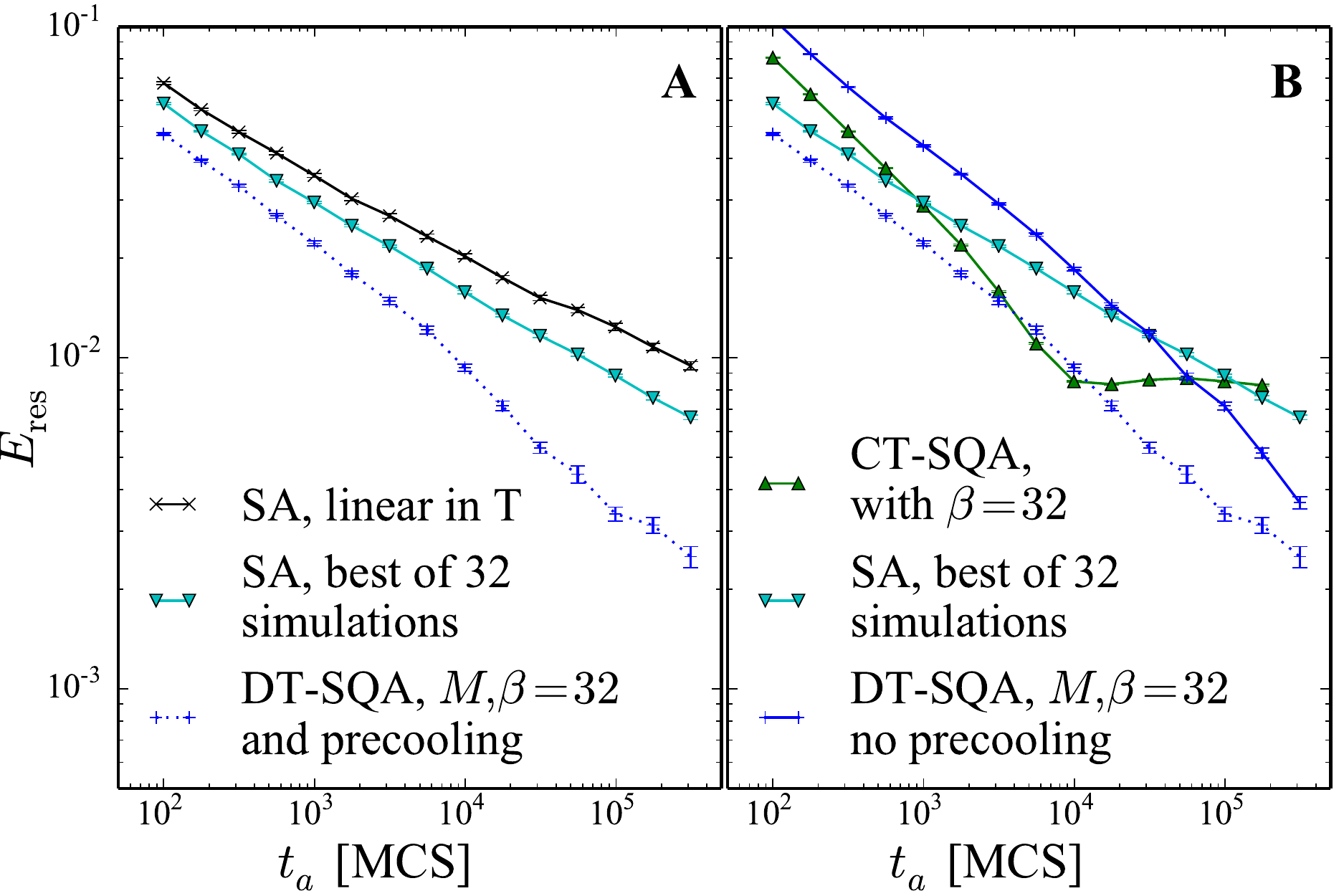}
  \caption{ Decrease of the residual energy $E_{\rm res}$ for SA, DT-SQA (panel A) and CT-SQA (panel B) as a function of the annealing time $t_a$ [in units of Monte Carlo steps (MCS)]  for the square lattice Ising spin glass instance of Ref. \protect\cite{santoro1} with 6400 spins and uniformly distributed couplings in (-2,2). The plotted value of $E_{\rm res}$ and error bars are obtained by averaging over forty annealing runs.}
  \label{fig:fig1}
\end{figure}

In order to investigate these seemingly contradictory results we first show, in Fig. \ref{fig:fig1}A, that we can reproduce the results of Ref. \cite{santoro1}. Additionally, we show the best results of 32 independent SA simulations, which corresponds to roughly the same 
computational effort as the SQA simulations, as illustrated in Fig. \ref{fig:path}C.  In either case we see that, as in Ref. \cite{santoro1}, the scaling of SQA is superior to that of SA.

However, these simulations were all performed with a finite number of time slices $M$ and a corresponding non-zero time step $\Delta_\tau=\beta/M=1$, which we refer to as a discrete time SQA  (DT-SQA) simulation.  Discrete time steps incur  time discretization errors of order  $\mathcal{O}( {\beta^3}/{M^2} )$. To obtain accurate  thermal averages for the quantum system one has to either extrapolate DT-SQA results to  $\Delta_\tau\rightarrow0$ or perform a continuous time SQA simulation (CT-SQA) that works directly in the limit $\Delta_\tau\rightarrow0$ \cite{ctq}. 

Repeating the same simulations using CT-SQA in Fig. \ref{fig:fig1}B we see that the CT-SQA result has an entirely different behavior than the corresponding DT-SQA curve.  While the performance is improved for fewer than $10^4$ Monte Carlo steps (MCS, corresponding to one attempted update per spin), the residual energy saturates for longer annealing times, at a level higher than that reached by SA. While the time discretization error in DT-SQA is of no concern for its use as a classical optimization algorithm \footnote{The time discretization error vanishes at the end of the annealing schedule when the transverse field is switched off and all remaining terms in the Hamiltonian commute}, it does not reflect our expectations for a physical quantum device, for which the continuous time limit is relevant. Hence, the circumstances under which SQA outperforms SA depend on whether we use SQA as a quantum inspired classical algorithm, or as simulation of a physical system. Understanding the role of time discretization in SQA is important both to estimate the performance of experimental quantum annealers as well as to tune SQA as a classical algorithm.

\begin{figure*}
  \centering
    \includegraphics[width=\columnwidth]{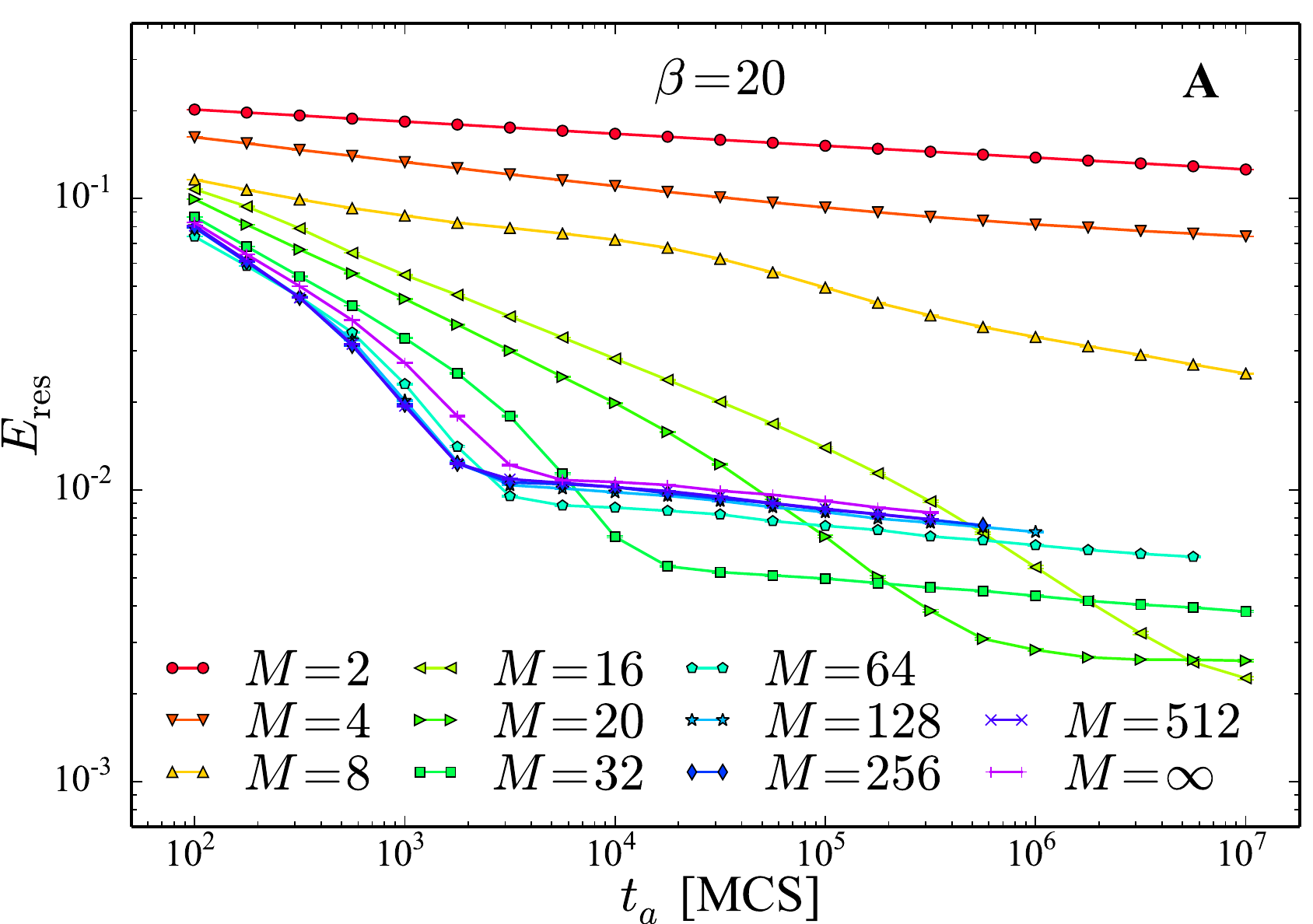} 
      \includegraphics[width=\columnwidth]{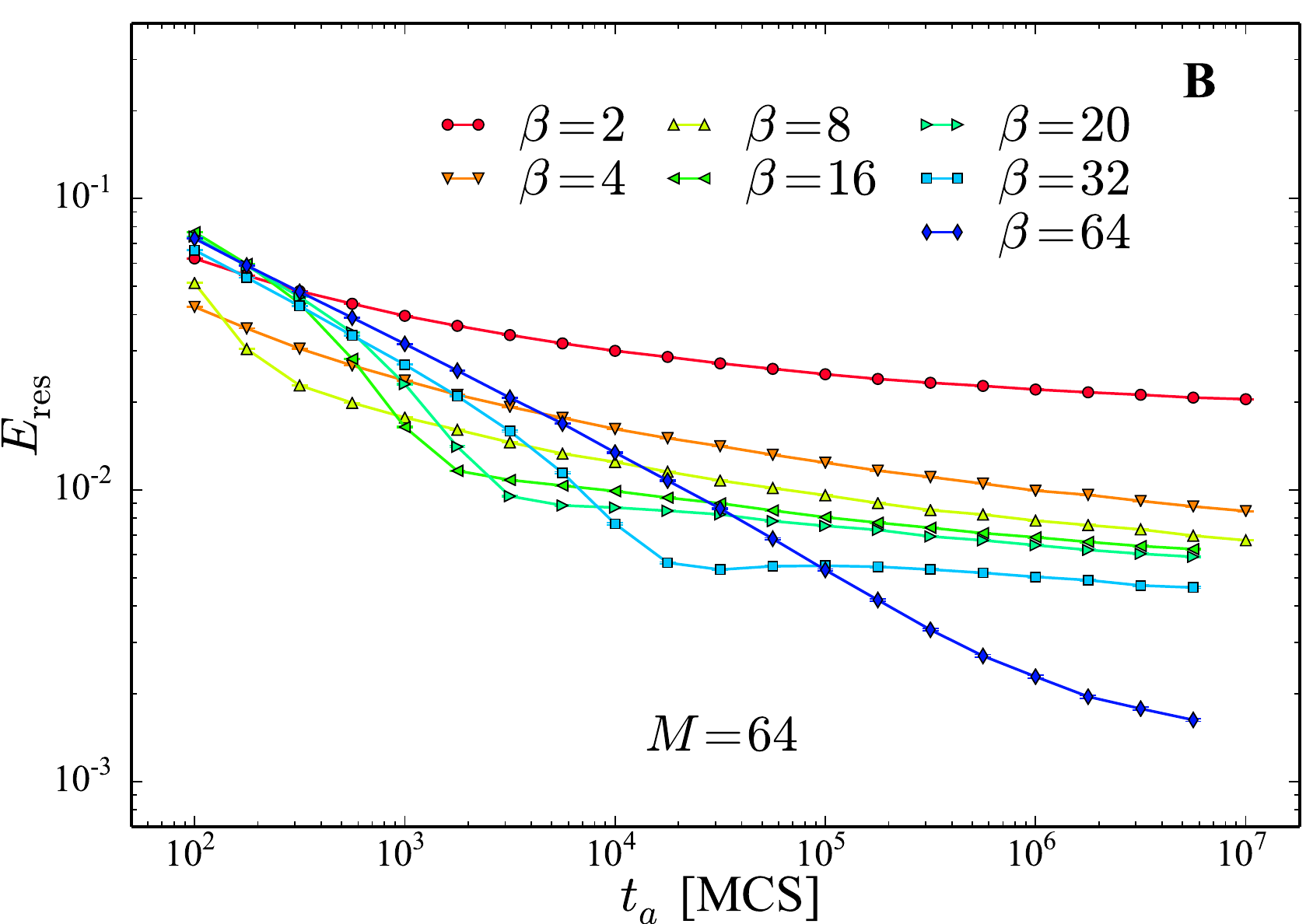}
    \caption{A) Convergence of DT-SQA towards the continuous time limit $M=\infty$ obtained by CT-SQA. Shows is the average of 1000 different disorder realizations  annealed  at an inverse temperature of $\beta=20$. The lowest energy configuration along the imaginary time axis was taken to calculate the residual energy. 
    B) Temperature dependence of  DT-SQA with a constant number of $M=64$ time slices. Lowering the temperature increases the time step $\Delta_\tau=\beta/M$. This reduces the initial drop in energy but allows to ultimately find a final configuration with lower energy. }
  \label{fig:fig2}
\end{figure*}

\paragraph{ Effects of Time Discretization and Temperature:}

To understand the role of time discretization we have gone beyond the single spin glass instance of Ref. \cite{santoro1} and studied 1000 random spin glass instances on an $80\times80$ square lattice with periodic boundary conditions. We use the  the same distribution, choosing the
 $J_{ij}$ uniformly from the interval $(-2,2)$, and set all local fields to zero ($h_i=0$) and obtain the exact ground state energy using the spin glass server. Following the procedure of Ref. \cite{santoro2}, the initial state in Fig. \ref{fig:fig1} was prepared by precooling. Comparing the DT-SQA curves with and without precooling (Fig. \ref{fig:fig1}B), we find that precooling only results in constant offset but does not improve the scaling. We thus omit precooling from our simulations.

In Fig. \ref{fig:fig2}A we show the residual energy as a function of
annealing time for various Trotter numbers
$M$. As expected, for $M\rightarrow \infty$, DT-SQA converges towards the continuous time limit. For the chosen temperature of $\beta=20$,
convergence is achieved for $M\ge128$. We find the same surprising behavior already indicated in Fig. \ref{fig:fig1}: Although the initial scaling is better in the continuous
limit, lower residual energies are reached at a finite time step size. Comparing $M=16$ and $M=64$, a lower residual energy of $2 \cdot 10^{-3}$ is found for $M=16$ compared to $5 \cdot 10^{-3}$ for $M=64$, despite the computational effort being four times smaller.

Analyzing the residual energies as a function of temperature with a 
constant number of time slices, as shown in Fig. \ref{fig:fig2}B, leads to a similar observation.
For $\beta < 20$, the DT-SQA results match well with the CT-SQA results (shown in the Supplementary Material), indicating that 64 time slices are sufficient to converge to the continuous time limit. At lower temperatures deviations from CT-SQA are seen and  the larger time step $\Delta_\tau=\beta/M$ in DT-SQA allows to eventually find states with lower energy than in CT-SQA -- consistent with the results of changing $M$. 
%%%%%%%%%%%%%%%%%%%
A closer look at Fig. \ref{fig:fig2}B shows that at lower temperatures lower energies can be reached. This fact, which is confirmed in the continuous time limit shown in the Supplementary Material, is encouraging for a potential weak quantum speedup \cite{article_eth} for SQA over SA in the zero-temperature limit. 

For all choices of $M$ and $\beta$ we considered, the residual energy saturates at some point, indicating that the simulations consistently get stuck in some local minimum during annealing. We will discuss reasons for this behavior below.

\paragraph{Quantum Annealing as a Classical Optimization Method: }
When discussing SQA as a classical optimization algorithm, we can search the final configuration for the time slice (or time interval in continuous time) with the lowest energy. This improves the results if the spin alignment along the imaginary time axis is incomplete at the end of the annealing. However, we
have to take into consideration the increased computational effort of QMC simulations compared to SA. The number of Monte Carlo steps needs to be multiplied by $M$ for DT-SQA and by $\beta$ for CT-SQA.

\begin{figure*}[t]
  \centering
 \includegraphics[width=1.12\columnwidth]{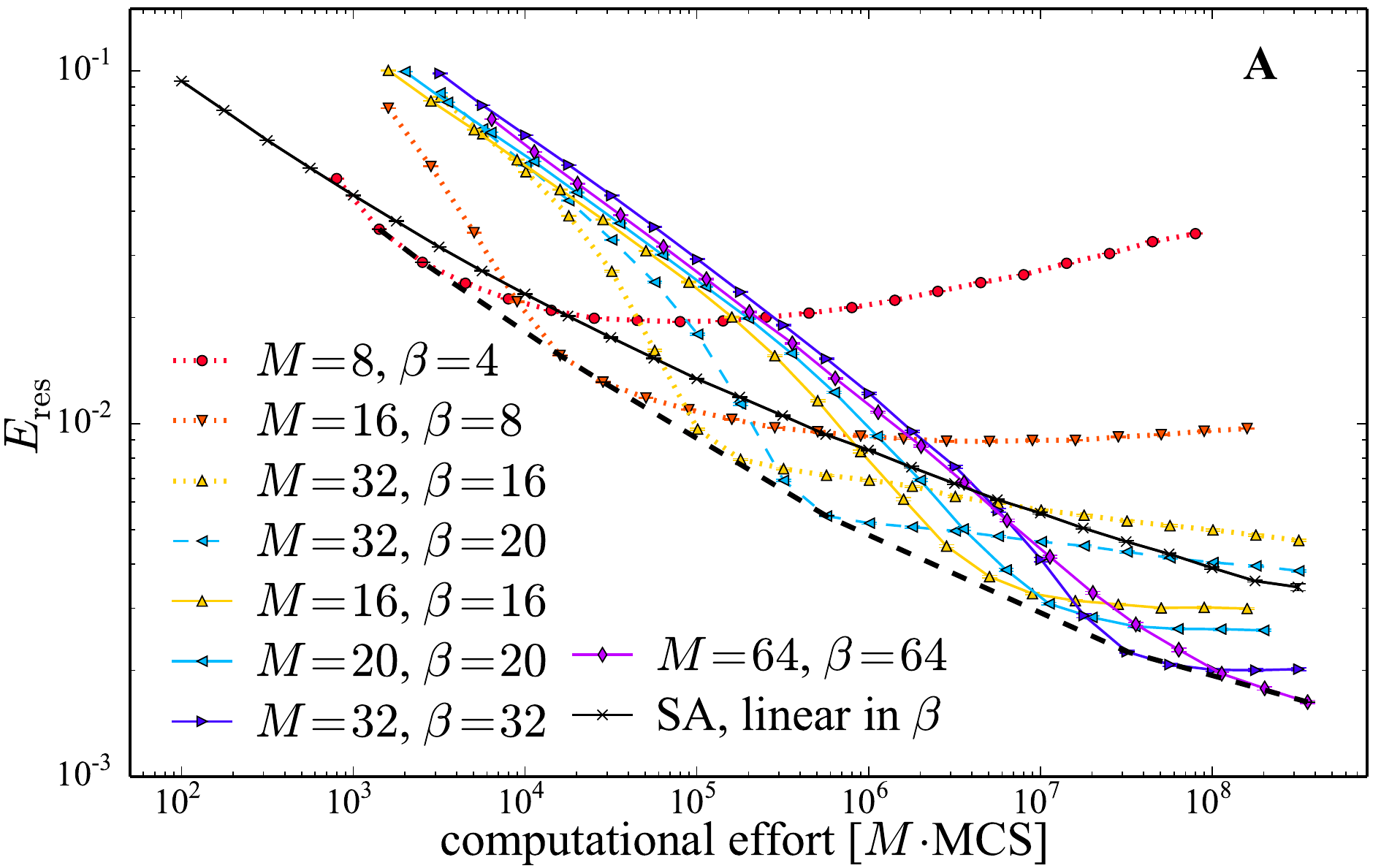}
 \includegraphics[width=0.88\columnwidth]{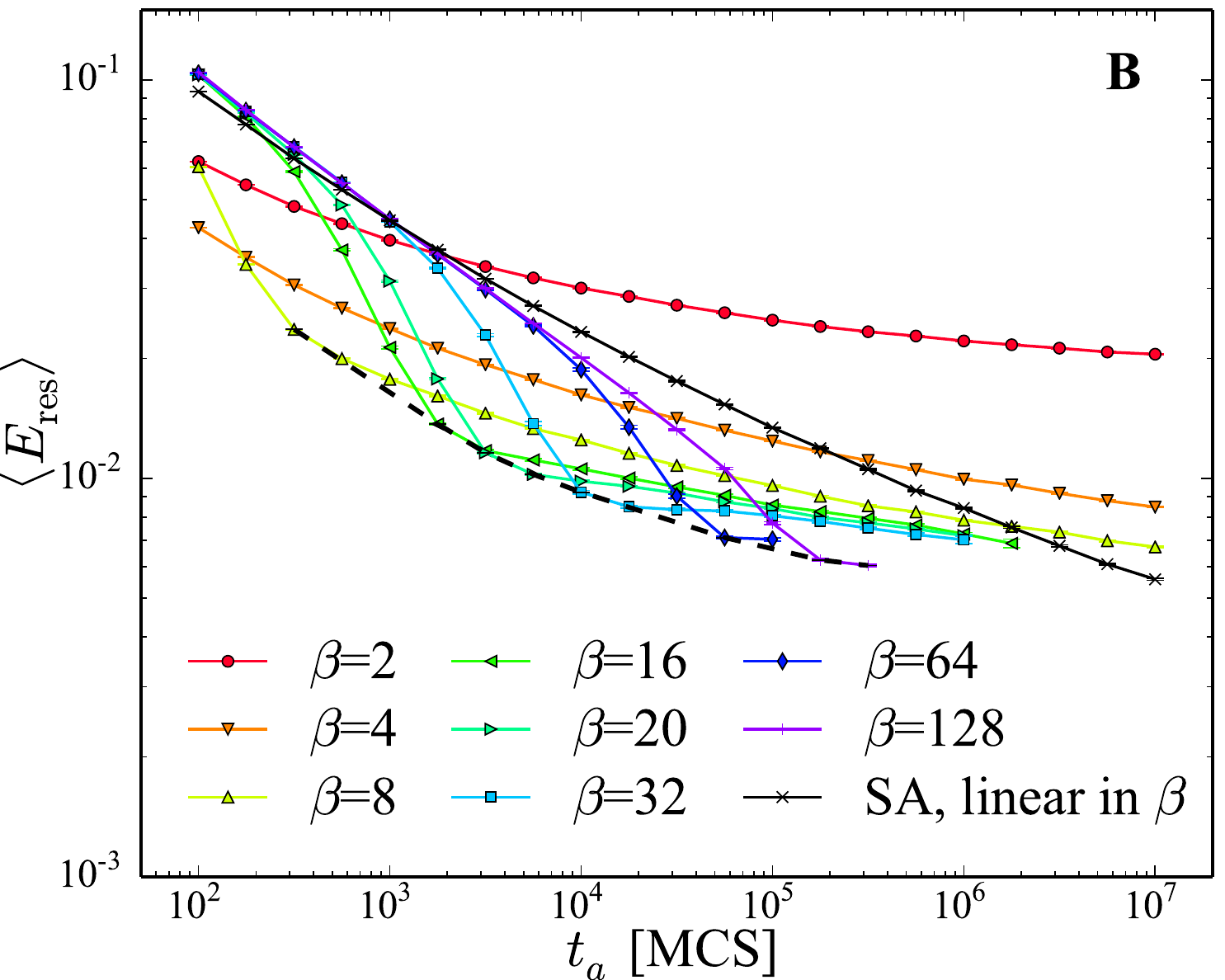}
   \caption{A) SQA as a classical optimizer: Choosing a suitable annealing temperature and time step DT-SQA scales better than SA, consistent with the results of Ref. \protect\cite{santoro1}.  B) SQA as a simulation of a  physical system: here QA is performed using DT-SQA with large enough $M$ to be converged to the continuous time limit. To be relevant for comparison to quantum devices we here average the final energy over imaginary time instead of picking the lowest energy configuration.}
  \label{fig:all_averages}
\end{figure*}

Plotting the residual energy as a function of total computational effort in Fig. \ref{fig:all_averages}A we find that -- in agreement with Ref. \cite{santoro1} -- with suitable chosen temperature and number of time steps, DT-SQA outperforms SA. The optimal choice depends on the desired computational effort and the envelope seems to outperform SA, although the asymptotic behavior when we anneal for longer times seems similar. 

In order to use SQA as a classical optimization algorithm it is thus advantageous to use a small time step $\Delta_\tau$ for short annealing times, since the continuous time limit has a more rapid initial decrease of $E_{\rm res}$. When annealing for longer times a lower temperature and larger time step $\Delta_\tau$ are preferred, since that way we reach lower asymptotic residual energies. To reach the lowest energies, rather large time steps of order unity are preferred, where the system consists of few moderately coupled individual replicas instead of a more tightly coupled continuous path of configurations. We note finally that, as we show in the supplementary material, even CT-SQA with suitably chosen temperature can outperform SA when used as a classical optimizer.

\paragraph{Quantum Annealing of a Physical System:}
While the Monte Carlo dynamics in SQA is not the same as the unitary or open systems dynamics of a physical quantum annealer, it is very similar since it captures tunneling and quantum entanglement. In particular, if  thermalization (at least within a local minimum) is fast compared to the annealing time, SQA  is expected to reliably capture the performance of physical QA, as has been seen in the case of the D-Wave devices \cite{Boixo:2014ej}.  

To use SQA as a tool to estimate the performance of hardware-based QA we have to take the continuous time limit and use either CT-SQA, or DT-SQA with a large enough number of time slices $M$ to be converged to the continuous time limit. We may measure only properties that are experimentally accessible and thus instead of picking the time slice with lowest energy, we either have to average the residual energy over all time slices, or measure it just at one randomly chosen imaginary time to mimic the process of measurement in a quantum system.

Figure \ref{fig:all_averages}B  shows the slightly higher residual energy obtained this way as a function of the number of MCS. We find that increasing the temperature slightly over that when SQA is used as a classical optimizer helps performance. For more details we refer to the Supplementary Material. Lower temperatures are again preferred for longer annealing times. While SQA outperforms SA for short annealing times, the asymptotic scaling of the envelope seems worse for SQA.

We find that for short annealing times, up to $t_a=10^5$ MCS, SQA still outperforms SA when choosing an appropriate temperature but the asymptotic scaling is better for SA.

\paragraph{Discussion and Outlook:}

Carefully investigating evidence for quantum annealing outperforming classical annealing for spin glass instances, we found that, surprisingly, the performance advantage previously observed for path-integral QMC annealing compared to classical annealing  \cite{santoro1,santoro2} is due to the use of large imaginary time steps in the path integral and choosing the lowest energy over all time slices. When taking the physical limit of continuous time and measuring the average energy, the advantage vanishes.

We also found that SQA tends to get stuck in local minima more than SA. This may be understood by the more deterministic dynamics of QA, preferring a subset of low-energy states over others \cite{Matsuda:2009ji}. Repeating SQA can thus get consistently stuck in similar local minima. SA, on the other hand, starts in a random state at high temperatures and thus explores the configuration space more evenly.
The more deterministic nature of SQA can also explain the counterintuitive result that for some choices of parameters (see Fig. \ref{fig:all_averages}) the residual energy may increase when annealing more slowly. As pointed out by Ref. \cite{Crosson:2014ez}, perturbing a quantum annealing schedule, for example by annealing faster, can excite a system out of a local minimum in which QA is stuck and thus help to ultimately find a better solution.

Our results also resolve the discrepancy between the observed superiority of SQA over SA \cite{santoro1,santoro2} and recent arguments that two-dimensional spin glasses should not see any quantum speedup in QA \cite{Katzgraber:2014cy}. It will be interesting to explore if three-dimensional spin glasses or spin glasses with long range couplings exhibit indications of superiority for QA. When investigating the powers of QA for such spin glasses or for problem instances derived from applications, it will be important to compare to both discrete and continuous time SQA. The former being relevant for the assessment of the powers of SQA as a classical optimization algorithm and the latter for evidence of potential quantum speedup on quantum annealing devices.

We thank H.G. Katzgraber, G. Santoro, and I.~Zintchenko for useful discussions,  G. Santoro for providing the spin glass instance used in Ref. \cite{santoro1}, and Canadian Maple Syrup for inspiration \cite{bet}. This work was supported by the Swiss National Science 
Foundation through the National Competence Center in Research QSIT and by the European Research Council through ERC Advanced Grant SIMCOFE. M.T. acknowledges the hospitality of the Aspen Center for Physics, supported by SNF grant 1066293. The spin glass server \cite{sgserver} was used to obtain the ground states for our problem instances.

\bibliography{biblio.bib}{}

\end{document}